\documentclass[a4paper,11pt]{article}
\pdfoutput=1 

\usepackage{jheppub} 

\usepackage[T1]{fontenc} 

\usepackage{caption}
\usepackage{subcaption}

\def\eps{\epsilon}

\newcommand \ket [1] {|{#1}\rangle}
\newcommand \bra [1] {\langle {#1}|}

\newcommand{\cM}{{\cal M}} 
 
\newcommand{\cS}{{\cal S}} 

\newcommand{\cL}{{\cal L}} 
\newcommand{\cF}{{\cal F}}

\newcommand{\cO}{{\cal O}}

\newcommand{\ep}{\epsilon}

\renewcommand{\a}{\alpha}

\newcommand{\nn}{\nonumber}
\newcommand{\Ga}[2]{\Gamma_{#1}^{(#2)}}
\newcommand{\web}[2]{w^{(#1,#2)}}
\newcommand{\com}[2]{\left[#1,#2\right]}
\newcommand{\T}{\mathbf{T}}

\newcommand{\Calum}[1]{\textcolor{magenta}{[#1]}}

\title{A new method for calculating the soft anomalous dimension matrix for massive particle scattering}
\author[a]{Johannes Henn,}
\author[b]{Calum Milloy,}
\author[c,d]{Kai Yan}

\affiliation[a]{Max-Planck-Institut f\"{u}r Physik, Werner-Heisenberg-Institut, D-80805 M\"{u}nchen, Germany}
\affiliation[b]{Dipartimento di Fisica and Arnold-Regge Center, Universit\`a di Torino, and INFN, Sezione di Torino, Via P. Giuria 1, I-10125 Torino, Italy}
\affiliation[c]{INPAC, Shanghai Key Laboratory for Particle Physics and Cosmology, School of Physics and Astronomy, Shanghai Jiao Tong University, Shanghai 200240, China}
\affiliation[d]{Key Laboratory for Particle Astrophysics and Cosmology (MOE), Shanghai 200240, China}

\emailAdd{henn@mpp.mpg.de}
\emailAdd{milloy@to.infn.it}
\emailAdd{yan.kai@sjtu.edu.cn}

\preprint{MPP-2023-242}

\abstract{ 
The general structure of infrared divergences in the scattering of massive particles is captured by the soft anomalous dimension matrix. 
The latter can be computed from a correlation function of multiple Wilson lines.
The state-of-the-art two-loop result has a tantalizingly simple structure that is not manifest in the calculations. 
We argue that the complexity in intermediate steps of the known calculations comes from spurious, regulator-dependent terms.
Based on this insight we propose a different infrared regulator that is associated to only one of the Wilson lines.
We demonstrate that this streamlines obtaining the two-loop result:
computing the required Feynman integrals via the differential equations method, 
only multiple polylogarithmic functions appear (to all orders in the dimensional regulator), as opposed to elliptic polylogarithms.
We show that the new method is promising for higher-loop applications 
by computing a three-loop diagram of genuine complexity, and provide the answer in terms of multiple polylogarithms.
The relatively simple symbol alphabet we obtain may be of interest for bootstrap approaches.
}

\begin{document}

\setcounter{tocdepth}{2}
\maketitle
\setcounter{page}{1}

\section{Introduction}

Being able to predict the structure of infrared divergences in scattering processes is relevant for collider physics.
For massless particles, the general structure of divergences is known to three loops. 
The infrared divergences factorize, similar to ultraviolet divergences. 
Due to color dependence, the infrared factorization takes the form of a matrix in color space. 
The essential part of the matrix is computed by the soft anomalous dimension matrix, which is determined from a correlation function of Wilson lines meeting at the origin, and pointing in the direction of the scattered particles' momenta \cite{Korchemskaya:1994qp}.

Up to two loops the soft anomalous dimension matrix consists only of so-called dipole terms, which depend on pairs of scattered particles only \cite{Dixon:2008gr,Becher:2009cu}. Starting from three loops, correlations between multiple lines (in particular, four lines) are also present. Interestingly, the four-line correlations depend on a function of two scale-invariant ratios, which has been computed at three loops \cite{Almelid:2015jia}, and its presence has been confirmed in four-particle scattering amplitudes \cite{Henn:2016jdu,Caola:2021rqz}.

The structure of infrared divergences of massive particles is more intricate.  In the two-line case, the state of the art for the cusp anomalous dimension is three-loop order in QCD \cite{Grozin:2014hna,Grozin:2015kna} and four-loop order in QED \cite{Bruser:2020bsh}.  Much less is known beyond the dipole terms. 
The two-loop soft anomalous dimension matrix involving three massive particles was computed by several groups \cite{Mitov:2009sv,Ferroglia:2009ep,Mitov:2010xw,Chien:2011wz}.
The result includes a color structure beyond the dipole terms, namely $f^{abc} {\bf T}_{i}^{a}{\bf T}_{j}^{b}{\bf T}_{k}^{c}$, 
where $\T_i^a$ is the colour matrix 
in the representation of particle $i$~\cite{Catani:1996vz,Bassetto:1983mvz}. (At higher loop orders, further color structure may appear.)
The accompanying function depends on the scale invariant combination of the scalar products between the velocities of the particles,
\begin{align}\label{eq:betavariables:introduction}
    \beta_{ij} = \frac{v_i\cdot v_j}{\sqrt{v_i^2}\sqrt{v_j^2}}\,.
\end{align}
The function multiplying this color structure depends on three variables of the type (\ref{eq:betavariables:introduction}), since at two loops Feynman diagrams connect up to three Wilson lines, cf. Fig.~\ref{fig:twoloopdiagrams}. Remarkably, the result is much simpler than a genuine three-variable function: it takes the form of a sum over products of logarithms and dilogarithms that individually depend on one variable only.  

This simplicity is not transparent in the state-of-the-art calculations.
It is interesting to note that the calculations in the literature were done by different means, via Mellin-Barnes methods \cite{Ferroglia:2009ep}, by a position-space calculation \cite{Mitov:2009sv,Mitov:2010xw}, and by a special gauge choice \cite{Chien:2011wz}. While all produce the correct result, it is not clear whether the methods can be easily generalized to the next loop order. Moreover, for the first two methods, the answer takes a simple form appears at the end of the calculation only.
This motivates us to understand why this is the case, and to look for other approaches.

In recent years, the differential equation method \cite{Kotikov:1990kg,Bern:1993kr,Gehrmann:1999as} has proven extremely useful for the calculation of Feynman integrals, especially thanks to new ideas about the transcendental structure of Feynman integrals \cite{Henn:2013pwa}. 
It is therefore interesting to see how the two-loop calculation would look like with this method. It turns out \cite{HennSimmonsDuffinUnbuplished,Milloy:2020hzi} that the most complicated two-loop Feynman diagram, shown in Fig.~\ref{fig:3gv}, involves elliptic functions.
This can be seen by a cut analysis, or by inspecting the differential equations matrix. 
What this means is that the $\mathcal{O}(1/\eps)$ part of the calculation, which contributes to the soft anomalous dimension matrix, is simple, but the higher order terms in the $\eps$-expansion are much more complicated. 
Although extracting the relevant information from the  complicated differential equations is possible \cite{Milloy:2020hzi}, by carefully expanding them in $\eps$, this is rather cumbersome. So, the question arises, is there a simpler approach?

It is important to realize that the soft anomalous dimension matrix corresponds to the leading divergence in $\eps$ of the Feynman integrals (or, linear combinations of Feynman diagrams organized in so-called webs \cite{Gatheral:1983cz,Frenkel:1984pz,Sterman:1981jc}).
In the calculations, a regularization choice is usually made to extract this divergence. In the language of position-space Wilson line correlators, the divergence we are interested in is a short-distance one, originating from webs of Feynman diagrams that homogeneously scale to zero towards the origin. This divergence is regulated by dimensional regularization, with $d=4-2\eps$, and $\eps>0$. The soft anomalous dimension matrix corresponds to the coefficient of the $1/\eps$ pole. However, the Wilson lines extend all the way to infinity, and this introduces a long-distance divergence that we are not interested in. It is common in the literature 
to regulate this by an exponential suppression factor. It is important to realize that due to the cutoff, only the leading $1/\eps$ pole is gauge-invariant and meaningful, while e.g. the finite term is not. So, it is no surprise that the functions contributing to the finite term can be very complicated - they do not have a physical meaning, and depend on the regularization procedure!

This motivates us to search for a different way of regulating the long-distance divergences of the webs. 
Our idea is the following:
to make the webs 
well-defined, it is sufficient to regulate one of the Wilson lines with an exponential suppression factor. 
In this paper, we provide an argument why this procedure leads to the correct soft anomalous dimension matrix.
We introduce the new method in section \ref{sec:method}.
We then show the effectiveness of the method in section \ref{sec:twoloop}, by easily reproducing the known two-loop result, and by computing the contribution of a previously unknown three-loop web function for the three-lines soft anomalous dimension matrix, in section \ref{sec:threeloop}. We discuss implications for prospects of future investigations in section \ref{sec:outlook}.

\section{Definitions and discussion of the method}
\label{sec:method}

 \subsection{Soft anomalous dimension matrix}

Scattering amplitudes of massive particles contain infrared divergences due to    
 virtual exchange of small quanta between heavy particles.  
In particular,  an
 $n$-point  gauge theory amplitude with massive external legs  obeys a factorization formula  which characterizes its all-loop infrared structure, 
\begin{align}
\mathcal{M}_n (  p_i, m_i, \mu   ;   \alpha_s(\mu) ; \epsilon ) =  S_n ( v_i ;  \alpha_s(\mu)  ;    \epsilon) 
 H_n (  p_i, m_i , \mu ;  \alpha_s(\mu) ) \,,
\end{align} 
where $S_n$ is the soft function encoding all the infrared singularities, and $H_n$ collects the finite remainder of the amplitude $\cM_n$. All quantities are renormalized with ultraviolet (UV) poles removed. 
 The running coupling $\alpha_s(\mu)$ is evaluated at the scale $\mu$ in $\overline{\rm MS}$ scheme. It satisfies the following equation in $d=4-2\eps$ dimensions,
 \begin{align}
     \frac{d  \ln \alpha_s}{ d  \ln \mu}  = -2 \epsilon  - 2 
 \sum_{k} \left( \frac{\alpha_s}{4 \pi} \right)^k b_k  \,.
  \end{align}
  The $b_k$ coefficients are coefficients of the beta function $\beta(\alpha_s)$.

The $S_n$ is a universal function depending on the velocities $v_i \equiv p_i/m_i$ of the scattering particles. It governs the singularity of the amplitudes  both in  the infrared region and in the infinite mass limit of the external particles,
 which can be studied in the context of Heavy Quark Effective Theory (HQET) (see, e.g.~\cite{manohar_wise_2000}). 
 In the limit $m_Q \rightarrow \infty$,  the heavy quark  traveling with velocity $v_i$ behaves like a classical source, 
 whose interactions with soft gluons are represented by a Wilson line 
 \begin{align} 
 Y_{i} (x) \equiv {\mathbf P}  \, \text{exp} \left[ i g \int_{0}^\infty ds \,v_i \cdot  A^a (x+ v_i s) T^a \right]  \,.
\end{align}
Hence $S_n$ admits a formal definition as the correlation function of $n$ semi-infinite Wilson lines emanating from the origin, whose renormalization group equation defines the  corresponding soft anomalous dimension,  
\begin{align} 
S_{n } = \bra{0} \prod_{i=1}^n \,  Y_{i} (0) \ket{0}_{\rm{ren.}}, \quad \; 
\frac{d}{d \ln \mu}   S_n  = - S_n \,  \mathbf{\Gamma}_n \,.
\end{align} 
The anomalous dimension $\mathbf{\Gamma}_n$ is a matrix operator in color space acting on the multi-point amplitudes. 
 It can be expanded as a sum over dipole, tripole and quadrupole functions, etc, 
 \begin{align} 
\mathbf{\Gamma}_{n} ( v_i ; \alpha_s ) = \sum_{(i,j)} \mathbf{\Gamma}_{ij} ( x_{ij};  \alpha_s )  + \sum_{(i,j,k)} \mathbf{\Gamma}_{ijk} (x_{ij}, x_{jk}, x_{ik}; \alpha_s)+ \cdots ,
\end{align}
which are parametrized by the variables 
 \begin{align}\label{eq:xDef}
     v_i \cdot v_j  \equiv  \cosh \phi_{ij} \equiv -\frac12 ( x_{ij} + 1/x_{ij}) \,,
 \end{align}
where we assume for simplicity $v_i^2=1$. The operators $\mathbf{\Gamma}_{ij \cdots}$ come from multi-Wilson-line diagrams describing the interactions between 
Wilson lines labeled by $i,j, \ldots$, and hence depend on the associated color charges, as well as on the angles between the respective lines.
The corresponding light-like soft anomalous dimension matrix (which is reached via $i\phi_{ij}{\text 's}  \rightarrow  \infty$) is known up to the three-loop order \cite{Almelid:2015jia}. 
In this case, beyond-dipole terms vanish at two loops and begin at three loops. This is in contrary to the massive case, where beyond-dipole terms are already present at two loops.

The soft function obeys the exponentiation theorem for Wilson-line matrix elements, whose logarithm defines a
web function free from sub-divergences in the infrared regime. More explicitly, it can be constructed through its $\alpha_s-$expansion, 
\begin{align}
W (\alpha_s (\mu) )  \equiv  &  \log\, \langle Y_1 \cdots Y_n  \rangle_{\text{ren.}} \\
 =  & \log \bigg\{ 1+ \sum_{k=1}^\infty \left( \frac{\alpha_s(\mu) }{4\pi} \right)^k S_n^{(k)}  \bigg\}  \\
 =  &  \sum_{k=1}^\infty \left( \frac{\alpha_s (\mu) }{4\pi} \right)^k W^{(k)} \,.
\end{align}
In the abelian theory, e.g. QED, the $L-$loop web function determines the $L-$loop soft anomalous dimension through its evolution equation
 $  \frac{d}{d \ln \mu}  W_{\rm QED}^{(L)} = - \Gamma^{(L)}_{\rm QED}$. In particular, in the cases where all fermions are treated as massive, the abelian web function is one-loop exact.  
In the non-abelian theory like QCD, higher-loop web functions contain  maximally non-abelian  color structures,   
 which, at a given loop order, 
 vanish in the abelian theory and cannot be decomposed onto an abelian product of  lower-loop color structures. 
More intuitively, the $L-$loop color factors can be associated with web diagrams containing  $L-1$  interation vertices among the virtual soft particles. 
These non-abelian contributions give rise to multi-line color correlations in the web function.

We are interested in the tripole function $\mathbf{\Gamma}_{ijk}$ in the soft anomalous dimension. Up to the three-loop order it has the following structure, 
\begin{align}\nn
  \mathbf{\Gamma}_{123}    =\,  &    
i f^{abc}\mathbf{T}_1^a \mathbf{T}_2^b \mathbf{T}_3^c    \,  F_1(x_{12}, x_{13}, x_{23}; \alpha_s)  \,   \\  
& +\Big\{  
   [\mathbf{T}_1^c, \mathbf{T}_1^a]  \,  [ \mathbf{T}_2^b , \mathbf{T}_2^c ] \,  \{\mathbf{T}_3^a, \mathbf{T}_3^b\}    F_2   (x_{12}, x_{13}, x_{23}; \alpha_s)  \, + \nonumber \\
  &  \hskip3cm + (1,2,3) \rightarrow (2,3,1)  + (1,2,3) \rightarrow (3,1,2)  \Big\}\,.
  \label{eq:Gamma123}
\end{align}
Here the $\alpha_s$-expansion of $F_1$ starts at the two-loop order, i.e. 
\begin{align}
F_1  =\sum_{k=2}^\infty \left( \frac{\alpha_s}{4\pi} \right)^{k} F_{1}^{(k)}\,,
\end{align}
 and that of $F_2$ starts at three-loop order, i.e.  
 \begin{align}
 F_2  =\sum_{k=3}^\infty \left(\frac{\alpha_s}{4\pi} \right)^{k} F_{2}^{(k)} \,.
 \end{align} 
 In the light-like limit, the all-order $F_1$ vanishes due to Bose symmetry, and $F_2^{(3)}$ reduces to a constant \cite{Almelid:2015jia}. 
In this work we introduce a new method to compute the full angle-dependent functions $F_1^{(3)}$ and $F_2^{(3)}$.

\subsection{Proposal of a new infrared regularization procedure}

In dimensional regularization, Wilson-line correlators do not depend on a physical scale. 
The soft function can be regularized by  introducing offshellness regulators $\delta_i$ on each Wilson line, which cut off the infrared (IR) divergences,  
\begin{align}
 \mathcal{S}_{\delta_1, \cdots, \delta_n}  \equiv  \langle 0 |  \prod_i  Y_{i, \delta_i} |0  \rangle,  \quad  Y_{i, \delta_i} \equiv {\mathbf P}_{\delta_i}\, \text{exp} \left[ ig  \int_{0}^\infty  ds\, v_i \cdot A^a (s v_i)  T^a \right] \,,
\end{align}
where the modified path-ordered exponential reads 
\begin{align} 
{\mathbf P}_{\delta} \,  \text{exp}\left[  \int  ds \right] \equiv  1 & + \int  ds \,  \theta(s) \,e^{i\delta s} 
+ \frac{1}{2!} \iint d s_1 d s_2 \,  \theta(s_1) \theta(s_2-s_1) \,e^{i\delta s_2}  \nn \\
&  +  \frac{1}{2!} \iint d s_1 d s_2 \,  \theta(s_2) \theta(s_1-s_2) \,e^{i\delta s_1} +  \cdots  \,.
\end{align}
The modified soft function contains only UV $\epsilon-$poles, which can be collected into a multiplicative renormalization $\mathcal{Z}-$factor for the Wilson-line operator such that 
\begin{align} \label{standardSoftPole} 
\cS_{\delta_1, \dots,  \delta_n} (v_i, \delta_i ; \alpha_s ; \epsilon ) =  \cS^{\text{fin.}}_{\delta_1, \cdots ,  \delta_n} (v_i, \delta_i /
\delta_n, \mu/\delta_n  ; \alpha_s(\mu); \epsilon) 
\,  \mathcal{Z}_{n}^{-1}  (v_i; \alpha_s (\mu); \epsilon ) \,,
\end{align}
where $\cS^{\text{fin.}}$ is finite at $\epsilon=0$. The factor $\mathcal{Z}_n$ is a matrix in color space satisfying the evolution equation 
\begin{align} \label{RGZn}
\frac{d}{d \ln \mu} \mathcal{Z}_n = - \mathcal{Z}_n \, \mathbf{\Gamma}_n \,.
\end{align} 
In this way the soft anomalous dimension matrix can be determined order-by-order in $\alpha_s$ by extracting the UV $\epsilon$-poles of the modified corrector $\cS_{\delta_1, \cdots, \delta_n}$. 

The matrix $\mathcal Z_n$ encodes the UV divergences and is therefore independent of the choice of IR regulators. We have the freedom to set the parameters $\delta_i$ to be any nonzero value.  
The standard choice is to consider $\cS_{1, \cdots, 1}$  where  all $\delta_i$'s are set to 1, see e.g. \cite{Grozin:2015kna}.  
In this work we will adopt a new approach by setting all but one of them to be zero, namely  to only introduce the regulator on the last external leg. Hence we have the advantage to compute 
simpler HQET integrals.  

In the following we will explain how to apply the new method to determine the soft anomalous dimension matrix. 
For simplicity we focus on the three-line case where we consider $\cS_{0,0,1}$ in the $\overline{\rm MS}$ scheme. 

Let us argue that $\cS_{0,0,1}$ has the same $\epsilon-$pole structures as the ratio between the standard three-line and two-line correlators, i.e. $[\cS_{1,1}]^{-1} \cS_{1,1,1}$, which implies that the finite remainder $\mathcal{S}_{ 0,0,1}^{\text{fin.}}(v_i, \mu; \alpha_s(\mu) )$ obeys
\begin{align} \label{RGS001}
 \left[ \mu \frac{\partial}{\partial \mu } + \beta(\alpha_s) \frac{\partial}{\partial \alpha_s} \right]
 \mathcal{S}_{ 0,0,1}^{\text{fin.}} = {\bf \Gamma_{2}} \,\mathcal{S}_{0,0, 1}^{\text{fin.}} -\mathcal{S}_{0,0, 1}^{\text{fin.}} \, {\bf{\Gamma_3} }  \,. 
\end{align}
To see why the statement is true, it is helpful to introduce modified correlators $\cS_{\delta, \delta }$ and 
$\cS_{\delta, \delta ,1}$ where $\delta \ll 1 $. 
Let us consider the ratio $[\cS_{\delta, \delta} (v_1, v_2)]^{-1} S_{\delta, \delta,1} (v_1, v_2, v_3) $.
Since the parameter $\delta$ is an IR cutoff, 
which characterizes the virtuality of the heavy particle $p^2 - m^2  \simeq m \delta$, 
the infrared singularity in $\cS_{\delta, \delta,1}$, 
as $\delta$ goes to zero, factorizes and cancels with the two-line correlator.  
Therefore the ratio $[\cS_{\delta, \delta}]^{-1} S_{\delta, \delta,1} $ is regular in the $\delta \rightarrow 0$ limit at fixed order in $\alpha_s$.  Formally we can take the limit $\delta \rightarrow 0$ in this ratio, yielding   
\begin{align} 
\lim_{\delta \rightarrow 0 }\;
 [\mathcal{S}_{ \delta, \delta } ]^{-1} \mathcal{S}_{ \delta, \delta, 1 } 
=  
  [\mathcal{S}_{ 0,0  } ]^{-1} \mathcal{S}_{ 0, 0, 1 }   =  
 \mathcal{S}_{ 0, 0, 1 }    \,, 
\end{align} 
where we set the scaleless function $\cS_{0,0} =1 $. Likewise,  
\begin{align} 
\lim_{\delta \rightarrow 0} \; [ S^{\rm{fin.}}_{\delta, \delta} ]^{-1}  S_{\delta, \delta ,1}^{\rm{fin.}} =  \mathcal{S}_{0,0,1}^{\rm{fin.}}  \,. 
\end{align} 
 Meanwhile, given the universal pole structure eq.~\eqref{standardSoftPole}, we have
\begin{align}
[\mathcal{Z}_2]^{-1} [S_{\delta, \delta}]^{-1}  \cS_{\delta, \delta, 1 } \mathcal Z_3  =  [ S^{\rm{fin.}}_{\delta, \delta} ]^{-1}  S_{\delta, \delta ,1}^{\rm{fin.}}    \,. 
\end{align} 
Taking the $\delta \rightarrow 0 $ limit on both sides  
we find 
\begin{align}\label{newSoftPole}
\mathcal{Z}_2^{-1}  \mathcal{S}_{0,0,1}  \mathcal{Z}_3 =   \cS_{0,0,1}^{\rm fin.} \,, 
\end{align} 
which is equivalent to the statement in eq.~\eqref{RGS001}. Using $\Gamma_2$, we can determine $\Gamma_3$ from $\mathcal{S}_{0,0,1}$ by demanding that the left-hand side of \eqref{newSoftPole} is finite.

The procedure works with any number of Wilson lines at any loop order, as we demonstrate presently. 
Indeed, writing $\Gamma_n=\sum_{i=1}^\infty\a_s^i\Gamma_n^{(i)}$,  and solving the renormalization group equation 
(\ref{RGZn}) gives
\begin{align}
\log \mathcal{Z}_n
=&\,
\a_s\frac{\Gamma_n^{(1)}}{2\ep}
+ \a_s^2\left(\frac{\Gamma_n^{(2)}}{4\ep}-\frac{b_0}{4\ep^2}\Gamma_n^{(1)}\right)
\\\nn
&+\a_s^3\left(\frac{\Gamma_n^{(3)}}{6\ep}
+
\frac1{48\ep^2}\left[\Gamma_n^{(1)},\Gamma_n^{(2)}\right]
-
\frac1{6\ep^2}\left(b_0\Gamma_n^{(2)}+b_1\Gamma_n^{(1)}\right)
+
\frac{b_0^2}{6\ep^3}\Gamma_n^{(1)}
\right)
\\\nn
&+
\a_s^4\bigg(
\frac{\Gamma_n^{(4)}}{8\ep}
+\frac1{48\ep^2}\left[\Gamma_n^{(1)},\Gamma_n^{(3)}\right]
-\frac{b_0}{8\ep^2}\Gamma_n^{(3)}
+\frac{\Gamma_n^{(2)}}{8\ep^2}\left(\frac{b_0^2}{\ep}-b_1\right)
\\\nn
&\hskip2cm-\frac{\Gamma_n^{(1)}}{8\ep^2}\left(\frac{b_0^3}{\ep^2}-\frac{2b_0b_1}{\ep}+b_2\right)
-\frac{b_0}{48\ep^3}\left[\Gamma_n^{(1)},\Gamma_n^{(2)}\right]
\bigg).
\end{align}
Next, we write 
\begin{align}\label{eq:wijDef}
\cS_{0,\dots,0,\delta}
=
\exp\left(
\sum_{i}\a_s^i w^{(i)}_n
\right)
=
\exp\left(
\sum_{i,j}\a_s^i\ep^{\,j} w^{(i,j)}_n
\right)\,, 
\end{align}
where $w_n^{(i)}$ is the coefficient of $\alpha_s^i$ in the exponent and $w_n^{(i,j)}$ is the coefficient of $\alpha_s^i\epsilon^{\,j}$. We have also reinstated an arbitrary regulator $\delta$. We then impose $\mathcal{Z}_{n-1}^{-1} \cS_n \mathcal{Z}_{n} = \text{finite}$, at each order in the coupling constant.   
At the first order this implies the relation
\begin{align}\label{eq:GammaN1}
 \Gamma_n^{(1)} =&\, \Gamma_{n-1}^{(1)} - 2 w_n^{(1,-1)}\,. 
\end{align}
Hence we obtain $\Gamma_n^{(1)}$ from $\Gamma_{n-1}^{(1)}$, supplemented by $w_n^{(1,-1)}$. 
Proceeding to the next orders, we have
\begin{align}
\Gamma_n^{(2)} =&\, 
\Gamma_{n-1}^{(2)} - 4 w_n^{(2,-1)} + 2 \left[\Gamma_{n-1}^{(1)},w_n^{(1,0)}\right]
-2\left[w_n^{(1,-1)},w_n^{(1,0)}\right]\,,
\label{eq:twoLoopSAD}
\\
\Ga{n}{3} =&
\Ga{n-1}{3}
-6 \web{3}{-1}_n
+\frac32 b_0 \com{\web{1}{-1}_n}{\web{1}{1}_n}
+
3 \com{\web{1}{0}_n}{\web{2}{-1}_n} 
\nn\\&
+ 3 \com{\web{2}{0}_n}{\web{1}{-1}_n} + 
 \com{\web{1}{0}_n}{\com{\web{1}{-1}_n}{\web{1}{0}_n}} - 
 \com{\web{1}{-1}_n}{\com{\web{1}{-1}_n}{\web{1}{1}_n}}
 \nn\\&+ 
 \frac32 \com{\Ga{n-1}{2}}{\web{1}{0}_n} - 
 \frac32 b_0 \com{\Ga{n-1}{1}}{\web{1}{1}_n} 
 \nn\\&+ 
 \frac34 \com{\Ga{n-1}{1}}{\com{\web{1}{-1}_n}{\web{1}{1}_n}} + 
 \frac34 \com{\web{1}{-1}_n}{\com{\Ga{n-1}{1}}{\web{1}{1}_n}} + 
 3 \com{\Ga{n-1}{1}}{\web{2}{0}_n} 
 \nn\\&- 
 \frac34 \com{\Ga{n-1}{1}}{\com{\Ga{n-1}{1}}{\web{1}{1}_n}} 
 + \frac32 \com{\web{1}{0}_n}{\com{\web{1}{0}_n}{\Ga{n-1}{1}}} \,.
\label{eq:threeLoopSAD}
\end{align}
Eqs.~\eqref{eq:GammaN1}, \eqref{eq:twoLoopSAD} and \eqref{eq:threeLoopSAD}  specify all the ingredients needed to determine the soft anomalous dimension matrix up to three-loop order. 
Requiring higher-order poles to cancel implies further relations. 
For instance we have
\begin{align}
   \label{eq:webConsistency}
   \web{2}{-2}_n
   =
   \frac14
  \com{\Ga{n-1}{1}}{\web{1}{-1}_n}\,.
\end{align}
This can be used as a consistency check when calculating the two-loop web function $w^{(2)}_n$.

We will demonstrate in the following section that eq.~\eqref{eq:twoLoopSAD} reproduces the correct two-loop result for ${\bf{\Gamma}}_3^{(2)}$.

\section{The two-loop soft anomalous dimension matrix via the new method}

\label{sec:twoloop}

\subsection{One loop}
\label{subsec:oneLoop}

The one-loop soft anomalous dimension can be found from the recurrence equation in eq.~(\ref{eq:GammaN1}). Solving eq.~(\ref{eq:GammaN1}) we find
\begin{align} \label{eq:Gn1Sum}
\Ga{n}{1} = -2 \sum_{i=1}^n \web{1}{-1}_i,
\end{align}
where $\web{1}{-1}_i$ is the single pole of the soft function that involves Wilson lines
$1$ to $i$, in eq.~(\ref{eq:wijDef}).
We recall that only the $n$-th Wilson line is IR-regulated.

\begin{figure}
     \centering
     \begin{subfigure}[b]{0.3\textwidth}
         \centering
         \includegraphics[width=\textwidth]{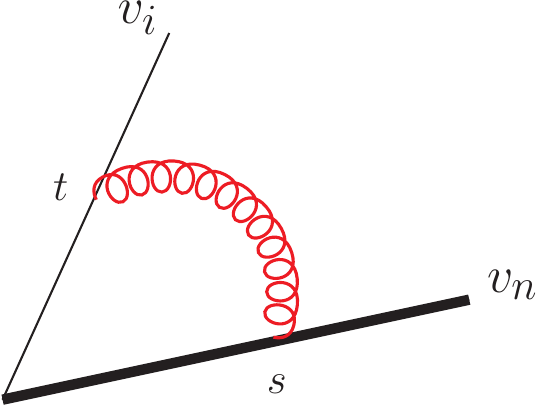}
         \caption{One-loop gluon exchange. \phantom{More text.}}
         \label{fig:oneLoop}
     \end{subfigure}
     \hfill
     \begin{subfigure}[b]{0.3\textwidth}
         \centering
         \includegraphics[width=\textwidth]{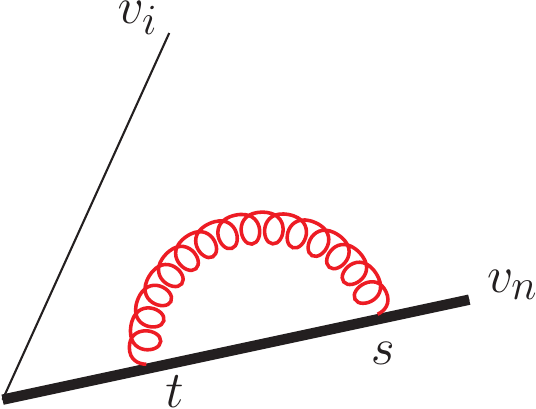}
         \caption{Self-energy correction on the regulated line.}
         \label{fig:SE}
     \end{subfigure}
     \hfill
     \begin{subfigure}[b]{0.3\textwidth}
         \centering
         \includegraphics[width=\textwidth]{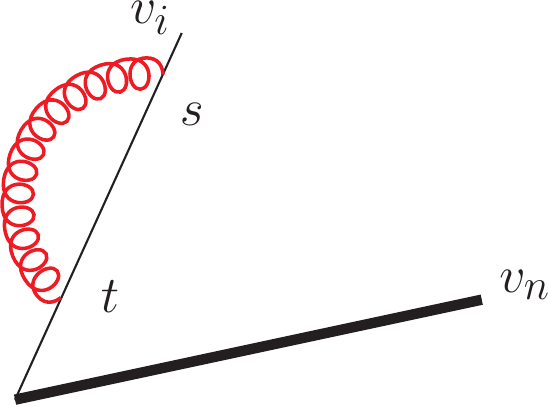}
         \caption{Self-energy correction on the unregulated line.}
         \label{fig:SEvansish}
     \end{subfigure}
        \caption{The three one-loop diagrams.}
        \label{fig:oneL}
\end{figure}

There are two contributions to $w^{(1)}_n$:
firstly, a gluon exchange between the regulated $n$-th Wilson line and an unregulated $i$-th line, with $i=\{0,\dots, n-1\}$, and secondly, the self-energy (SE) graph involving the $n$-th Wilson line. 
The former is shown in Figure~\ref{fig:oneLoop}, whereas the latter in Figure~\ref{fig:SE}.
Other potential contributions, due to gluon exchanges involving unregulated lines only, vanish as the integrals are scaleless. For example, the self-energy on an unregulated line is shown in Figure~\ref{fig:SEvansish}.
Note that this involves a cancellation between the IR and the UV.

It is convenient to compute the one-loop exchanges in configuration space where the Feynman-gauge gluon propagator is given by
\begin{align}
    \int\frac{d^dk}{(2\pi)^d}\frac{e^{-ik\cdot x}(-i g_{\mu\nu})}{k^2+i\varepsilon}
    =
    -\frac{\Gamma(1-\eps)}{4\pi^{2-\eps}}
    \frac{g_{\mu\nu}}{(-x^2 + i\varepsilon)^{1-\eps}}\,,
\end{align}
and where $x$ is the spacetime distance associated to the gluon propagator. 
In the diagram shown in
Figure~\ref{fig:oneLoop}, it is given by $x=sv_n-tv_i$. Here $s,t$ are line parameters that are integrated along the semi-infinite Wilson line segments, i.e. $s,t \in \lbrack 0,\infty\rbrack$. In Figure~\ref{fig:SE} we have $t\in \lbrack 0,s \rbrack$ and $s\in \lbrack 0,\infty \rbrack$.

We write the total contribution to $w_n^{(1)}$ as
\begin{align}
\a_s w_n^{(1)} =&\, 
g_s^2\mu^{2\eps}\frac{\Gamma(1-\ep)}{4\pi^{2-\ep}}
\bigg[
\sum_{i=1}^{n-1}\T_i\cdot\T_n\,
v_i\cdot v_n\int_0^\infty dsdt
\frac{e^{i\delta s}}{(-(s v_n-t v_i)^2+i\varepsilon)^{1-\eps}}
\nn\\&+
\frac{C_n}{2}\,
\int_0^\infty ds e^{i\delta s}\int_0^s dt
\frac{1}{(-(s v_n-t v_n)^2+i\varepsilon)^{1-\eps}}
\bigg]\,,
\label{eq:Wn1Int}
\end{align}
where $C_i = \T_i^a\T_i^a$ is the quadratic Casimir associated to line $i$.
The integrals in eq.~(\ref{eq:Wn1Int}) can be performed to all orders in $\eps$, resulting in a $_2F_1$ function. However, we will only require the expansion in $\eps$. In order to display the results, it is convenient to define the integral $f(x,\ep)$ and its expansion in $\ep$ as follows
\begin{subequations}
\begin{align}
    \label{def:fIntegral}
    f(x,\eps)
    =&\,
    -\frac12\,\left(x+\frac1{x}\right)
    \int_0^\infty dt \, \frac{1}{(t+\frac1{x})^{1-\eps}(t+x)^{1-\eps}}
    =
    \frac{r(x)}{2}\sum_{n=0}^\infty \frac{\eps^n}{n!}M_{n00}(x),
    \\
    \label{def:r}
    r(x)=&\,\frac{1+x^2}{1-x^2},
    \\
    \label{def:M000}
  M_{000}(x) =&\, 2\log(x),
  \\
  \label{def:M100}
  M_{100}(x) =&\, 2\text{Li}_2(x^2) + 4 \log(x)\log(1-x^2) - 2\log^2(x) - 2\zeta_2. 
\end{align}
\end{subequations}
The functions $M_{n00}$ are a subset of a larger class of functions $M_{klm}$ which form a basis of functions for multiple-gluon exchange webs 
\cite{Falcioni:2014pka}, where three- and four-gluon vertices are absent. We only display the explicit results up to $M_{100}$ as this is all we require to two loops. The one-loop web contribution in eq.~(\ref{eq:Wn1Int}) can be written as
\begin{align}
\a_s w_n^{(1)} =&\,
-\frac{\a_s}{\pi}\left(\frac{\mu^2}{\delta^2}\right)^{\eps}\frac{\Gamma(1-\ep)}{\pi^{-\ep}}
\bigg[
\sum_{i=1}^{n-1}\T_i\cdot\T_n\,
f(x_{in}, \eps)\Gamma(2\eps)
+
\frac{C_n}{2}
\Gamma(-1 + 2 \eps)
\bigg]
\\
=&\,
-\frac{\a_s}{\pi}\left(\frac{\bar\mu^2}{\overline{ \delta}^2}\right)^{\eps}e^{\eps\gamma_E}\Gamma(1-\ep)
\bigg[
\sum_{i=1}^{n-1}\T_i\cdot\T_n\,
f(x_{in}, \eps)\Gamma(2\eps)
+
\frac{C_n}{2}
\Gamma(-1 + 2 \eps)
\bigg],
\end{align}
where $\bar\mu^2=4\pi e^{-\gamma_E}\mu^2$ is the $\overline{\text{MS}}$ scale, $\overline{\delta} = 2\delta$ is a modified regulator and we have used the convenient $x_{ij}$ variables defined in eq.~(\ref{eq:xDef}).
The coefficients as an expansion in $\eps$ are then
\begin{align}\label{wn11}
\a_s w_n^{(1,-1)}
=&
-\frac{\a_s}{4\pi}
\left[
\sum_{i=1}^{n-1}\T_i\cdot\T_nr(x_{in})M_{000}(x_{in})
-C_n
\right],
\\ \label{wn10}
\a_s w_n^{(1,0)}
=&
-\frac{\a_s}{4\pi}
\left[
\sum_{i=1}^{n-1}\T_i\cdot\T_nr(x_{in})M_{100}(x_{in})-2\,C_n
\right]
\nn\\
&-
\frac{\a_s}{4\pi}
\log\left(\frac{\bar\mu^2}{\overline{\delta}^2}\right)
\left[
\sum_{i=1}^{n-1}\T_i\cdot\T_nr(x_{in})M_{000}(x_{in})
-C_n
\right].
\end{align}
Then using eq.~(\ref{wn11}) in eq.~(\ref{eq:Gn1Sum}), we have the result for the one-loop anomalous dimension
\begin{align}\label{eq:Gan1}
    \a_s\,\Ga{n}{1}
    =  \frac{\a_s}{2\pi}
    \sum_{i=1}^n
    \left[\sum_{j=1}^{i-1}\T_i\cdot\T_j\,
    r(x_{ij})M_{000}(x_{ij})
    -\,C_i
    \right].
\end{align}
Eq.~(\ref{eq:Gan1}) is in agreement with the classic result of Korchemsky and Radyushkin~\cite{Korchemsky:1987wg}. In our computation of $\Ga{n}{1}$, we have had to handle conceptual issues around extra infrared poles, as correlations between unregulated Wilson lines vanish. These are then accounted for in eq.~(\ref{eq:Gn1Sum}) by effectively adding them back. We will see in the remaining sections that using our regularisation scheme is combinatorically more complex than when all the lines are regulated due to a lack of symmetry. However, the integrals become simpler and can be computed using standard methods, as only one line is regulated.

\begin{figure}[t]
\centering
\begin{subfigure}[b]{0.45\textwidth}
\centering
  \includegraphics[width=\textwidth]{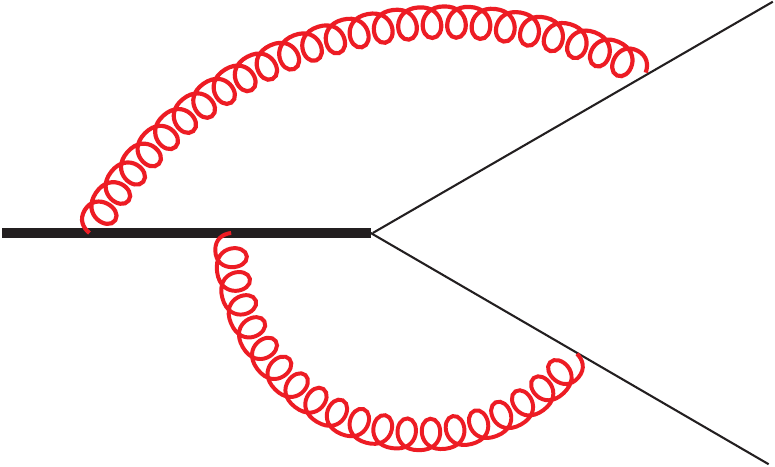}
  \caption{\label{fig:121b} }
\end{subfigure}
\begin{subfigure}[b]{0.45\textwidth}
\centering
  \includegraphics[width=\textwidth]{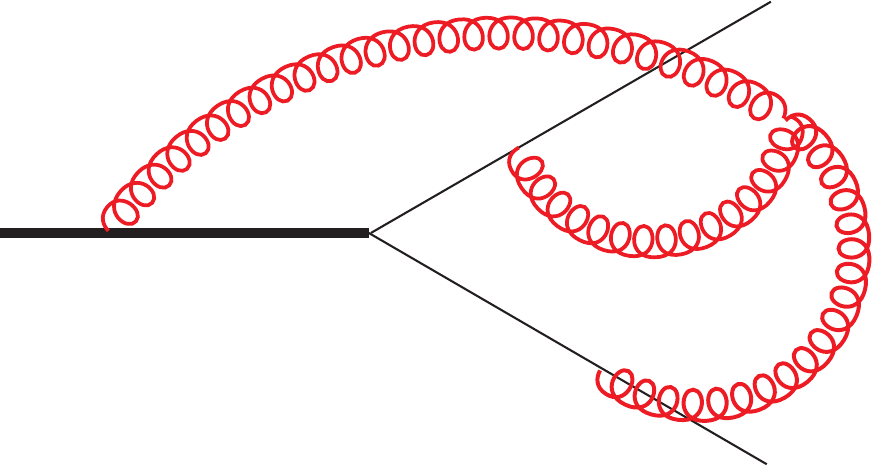}
  \caption{\label{fig:3gv} }
\end{subfigure}
\caption{\label{fig:twoloopdiagrams}
 Contributions to the two-loop soft anomalous dimension matrix. }
\end{figure}

\subsection{Two loops}
\label{subsec:twoloop}

The two-loop contribution to the soft anomalous dimension is described by eq.~(\ref{eq:twoLoopSAD}). 
In this section we compute it for the case of three Wilson lines, focusing 
on structures connecting all three lines, cf. eq.~(\ref{eq:Gamma123}).
In this case, we do not have the first term of eq.~(\ref{eq:twoLoopSAD}), $\Gamma_{2}^{(2)}$, and we do not consider any two-particle colour structures. 
Thus we write
\begin{equation}\label{eq:twoLoopAllThree}
    \Gamma_3^{(2)}\bigg\rvert_{{\text{
    connected}}}
    = \left\{
    - 4 w_3^{(2,-1)} + 2 \left[\Gamma_{2}^{(1)},w_3^{(1,0)}\right]
-2\left[w_3^{(1,-1)},w_3^{(1,0)}\right]
\right\}
\bigg\rvert_{\text{
connected}}.
\end{equation}
At two loops, there are two types of graphs connecting all three lines. One of them is the gluon exchange diagram shown in Figure~\ref{fig:121b}, which we label as $[1,2,1]$. (This notation keeps track of the number of gluon attachments to each line.)
The other graph is the three-gluon-vertex diagram shown in Figure~\ref{fig:3gv}. 
The calculation of these graphs is instructive.
The calculation of the gluon exchange diagram shows us how to deal with a subtlety related to infrared divergences, 
while that of the three-gluon-vertex diagram shows us the computational advantages of regulating only one Wilson line.

\subsubsection{Gluon exchange diagram}
\label{sec:121}

\begin{figure}
\centering
\begin{subfigure}[b]{0.45\textwidth}
\centering
  \includegraphics[width=0.8\textwidth]{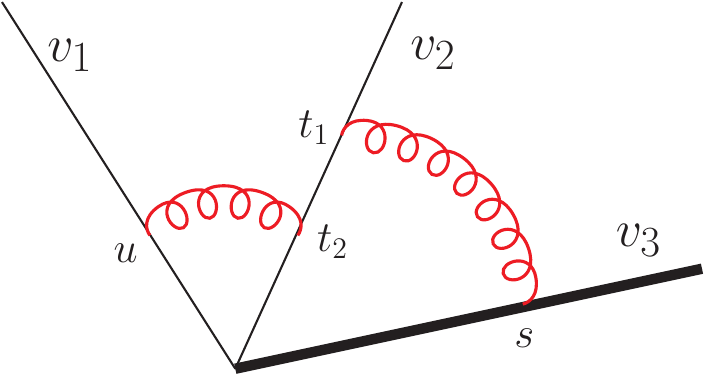}
  \caption{\label{fig:diaA} }
\end{subfigure}
\begin{subfigure}[b]{0.45\textwidth}
\centering
  \includegraphics[width=0.8\textwidth]{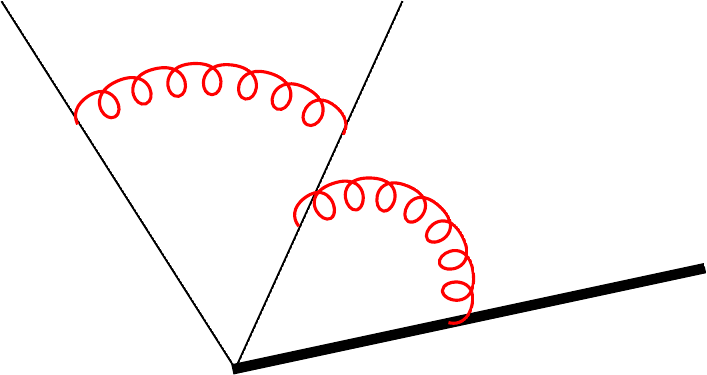}
  \caption{\label{fig:diaB} }
\end{subfigure}
\\
\centering
\begin{subfigure}[b]{0.45\textwidth}
\centering
  \includegraphics[width=0.8\textwidth]{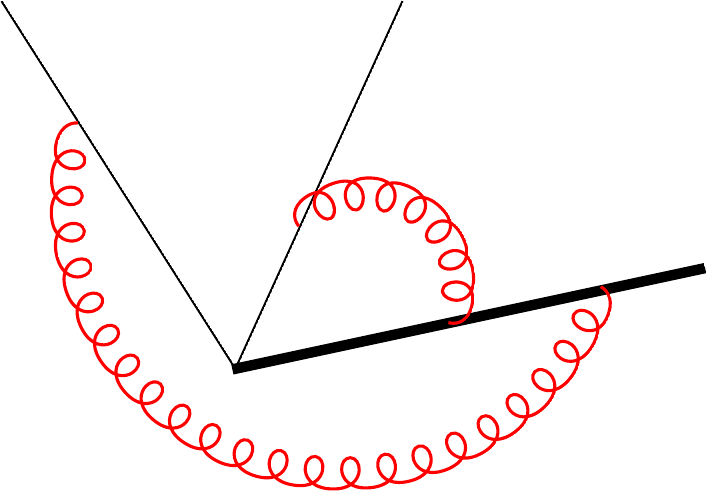}
  \caption{\label{fig:diaC} }
\end{subfigure}
\begin{subfigure}[b]{0.45\textwidth}
\centering
  \includegraphics[width=0.8\textwidth]{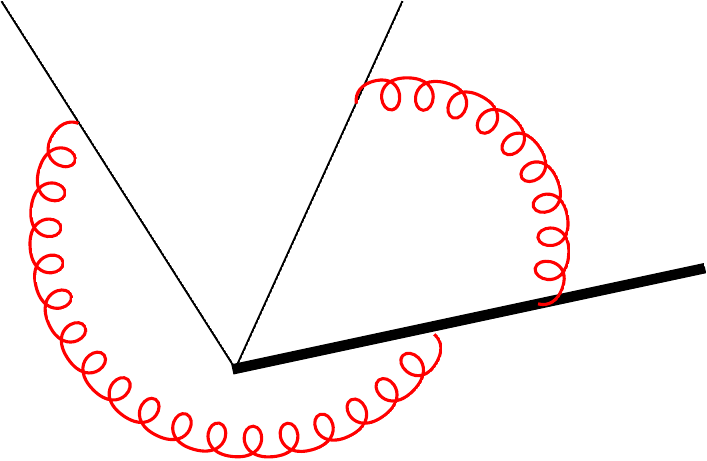}
  \caption{\label{fig:diaD} }
\end{subfigure}
\\
\centering
\begin{subfigure}[b]{0.45\textwidth}
\centering
  \includegraphics[width=0.8\textwidth]{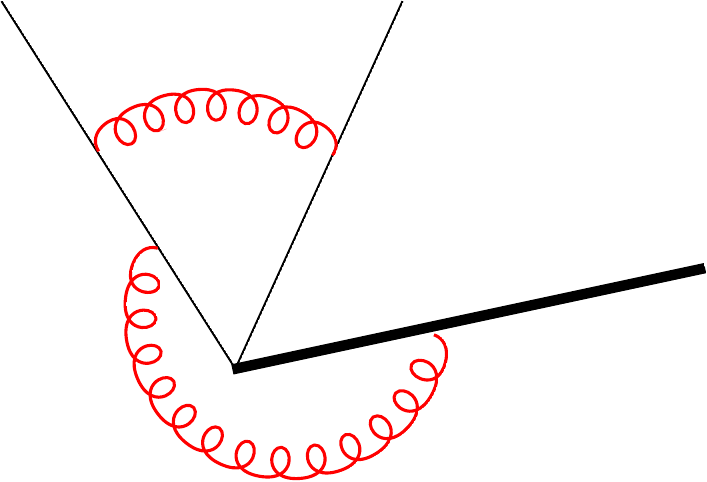}
  \caption{\label{fig:diaE} }
\end{subfigure}
\begin{subfigure}[b]{0.45\textwidth}
\centering
  \includegraphics[width=0.8\textwidth]{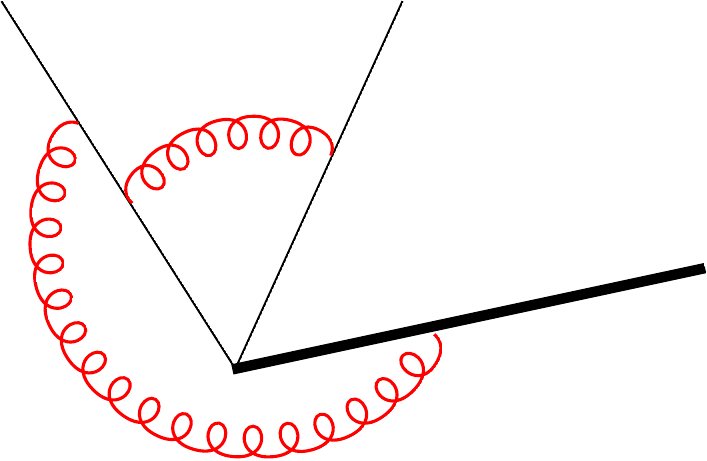}
  \caption{\label{fig:diaF} }
\end{subfigure}
\caption{\label{fig:121comp} Two-loop gluon exchange diagrams contributing to three-line color structures.}
\end{figure}

For the case of the $[1,2,1]$, the contributions to $w_3^{(2,-1)}$ in eq.~(\ref{eq:twoLoopAllThree}) are shown  
in Figure~\ref{fig:121comp}. The diagrams correspond to each permutation of gluon attachments to the Wilson lines. The thick Wilson line, $v_3$ is the regulated one. Diagrams 
\ref{fig:121comp}(f) and \ref{fig:121comp}(e) can be obtained from diagrams \ref{fig:121comp}(a) and \ref{fig:121comp}(b), respectively,
by interchanging $v_1$ and $v_2$. Therefore, we only need to compute diagrams \ref{fig:121comp}(a) - (d). 

Diagrams \ref{fig:121comp}(a) and \ref{fig:121comp}(b) have the kinematic factor
\begin{equation}
    F_{\ref{fig:121comp}(a)} - F_{\ref{fig:121comp}(b)} = (v_1\cdot v_2)(v_2\cdot v_3)\int_0^\infty du dt_1 dt_2 ds \, e^{i\delta s}\frac{\theta(t_1-t_2)-\theta(t_2-t_1)}{(-(sv_3-t_1v_2)^2)^{1-\eps}(-(uv_1-t_2v_2)^2)^{1-\eps}}\,, 
\end{equation}
whereas diagrams {\ref{fig:121comp}(c)} and {\ref{fig:121comp}(d)} are
\begin{equation}
    F_{\ref{fig:121comp}(c)} - F_{\ref{fig:121comp}(d)}
    =
    (v_1\cdot v_3)(v_2\cdot v_3)\int_0^\infty du dt ds_1 ds_2\frac{e^{i\delta s_1}\theta(s_1-s_2)-e^{i\delta s_2}\theta(s_2-s_1)}{(-(s_1v_3-uv_1)^2)^{1-\eps}(-(s_2v_3-tv_2)^2)^{1-\eps}},
\end{equation}
where the relative minus signs comes from non-Abelian exponentiation. It is reminded that the exponential damping factor $e^{i\delta s}$ is only added to the outer most attachment on $v_3$. The contribution to $w_3^{(2)}$ is
\begin{equation}
    w^{(2)}_{3,\,[1,2,1]}
    =
     \frac{   \mathcal{T}_{123}}{2\pi^2}
    \left[
    (F_{\ref{fig:121comp}(a)} - F_{\ref{fig:121comp}(b)}) - (F_{\ref{fig:121comp}(a)} - F_{\ref{fig:121comp}(b)})(v_1\leftrightarrow v_2) + F_{\ref{fig:121comp}(c)} - F_{\ref{fig:121comp}(d)}
    \right].
\end{equation}
Here we denoted the overall tripole colour factor by $\mathcal{T}_{123}\equiv if^{abc}\T_1^a\T_2^b\T_3^c$. 
The combination $F_{\ref{fig:121comp}(a)}-F_{\ref{fig:121comp}(b)}$ can be computed to give
\begin{align}
    F_{\ref{fig:121comp}(a)} - F_{\ref{fig:121comp}(b)} =&\,
    e^{-4\eps\gamma_E}(\delta^2)^{-2\eps}\frac{r(x_{12})r(x_{23})}{16}
    \bigg[
    \frac1{\eps^2}M_{000}(x_{12})M_{000}(x_{23})
    \\\nn&\hskip0.6cm+
    \frac1{\eps}\left(M_{000}(x_{12})M_{100}(x_{23})+M_{100}(x_{12})M_{000}(x_{23})\right)
    \bigg],
\end{align}
whereas
\begin{equation}
    F_{\ref{fig:121comp}(c)} - F_{\ref{fig:121comp}(d)} =0,,
\end{equation}
due to symmetry.
We define a modified $M_{100}$ function to absorb regulator-dependent terms as follows,
\begin{equation}
    M_{100}^{(\kappa)}(x) = M_{100}(x) + \log(e^{-\kappa\gamma_E}\delta^{-2\kappa})M_{000}(x).
\end{equation}
Using $w^{(1,-1)}$ from eq.~(\ref{wn11}), $w^{(1,0)}$ from eq.~(\ref{wn10}) and $\Gamma_2^{(1)}$ from eq.~(\ref{eq:Gan1}), the three contributions to eq.~(\ref{eq:twoLoopAllThree}) are then written as
\begin{align}
    -4\, w_{3,[1,2,1]}^{(2,-1)} =&\, \frac{\mathcal{T}_{123}}{8\pi^2}r(x_{12})\bigg\{r(x_{13})
    \left[M_{100}(x_{12})M_{000}(x_{13})+M_{000}(x_{12})M_{100}^{(2)}(x_{13})\right]
    \nn\\
    &\hskip2cm-r(x_{23})
    \left[M_{100}(x_{12})M_{000}(x_{23})+M_{000}(x_{12})M_{100}^{(2)}(x_{23})\right]
    \bigg\},
    \nn\\
    2\left[\Gamma_2^{(1)},w_{3}^{(1,0)}\right]
    =&\, \frac{\mathcal{T}_{123}}{4\pi^2}r(x_{12})M_{000}(x_{12})
    \left[r(x_{23})M_{100}^{(1)}(x_{23})-r(x_{13})M_{100}^{(1)}(x_{13})\right],
    \nn\\
    -2\left[w_{3}^{(1,-1)},w_{3}^{(1,0)}\right]
    =&\, \frac{\mathcal{T}_{123}}{8\pi^2}r(x_{13})r(x_{23})
    \left[M_{100}^{(1)}(x_{13})M_{000}(x_{23})-M_{000}(x_{13})M_{100}^{(1)}(x_{23})\right].
\end{align}
The sum of the above gives
\begin{equation}
    \Gamma_{3,[1,2,1]}^{(2)}
    =
    \mathcal{T}_{123}\left(\frac{1}{8\pi^2}\right)\sum_{i,j,k}\varepsilon_{ijk}\,r(x_{ij})\,r(x_{ik})\,M_{100}(x_{ij})\,M_{000}(x_{ik}),\label{eq:Gamma121}
\end{equation}
where all the dependence on the regulator $\delta$ cancels, as expected.

\subsubsection{The three-gluon-vertex diagram}
\label{sec:3gv}

To complete the calculation of the three-line soft anomalous dimension at two loops we require the coefficient of the single pole of Figure~\ref{fig:3gv}. 
Using our new method that places a regulator only on one Wilson line, 
we find that only four basis integrals are required in the top sector, as opposed to six in the traditional approach \cite{HennSimmonsDuffinUnbuplished,Milloy:2020hzi}. (See also \cite{Waelkens:2017zwt} for a calculation using unitarity cuts.) 

The contribution to the exponent of the soft function is written as
\begin{align}\label{def:w3gvDef}
    w^{(2)}_{3,\, 3gv}
    =&\,
    \frac{\mathcal{T}_{123}}{16\pi^2}\left(\bar\mu^2\right)^{2\eps}
    e^{2\eps\gamma_E}
    \nn\\
    &\times\int\frac{d^dk_1}{i\pi^{d/2}}\frac{d^dk_2}{i\pi^{d/2}}
    \int d^dk_3
    \frac{\delta^{(d)}(k_1+k_2+k_3)\sum_{i,j,k}\ep_{ijk}\left(v_i\cdot v_j\right)\left(k_i\cdot v_k\right)}{k_1^2k_2^2(k_1+k_2)^2(k_1\cdot v_1)(k_2\cdot v_2)(k_3\cdot v_3 - \delta)}.
\end{align}
We will conveniently set $\delta=1$ as the dependence on $\delta$ can be easily recovered by counting mass dimensions.
We define the integral family as
\begin{align}
I_{a_1a_2a_3a_4a_5a_6a_7a_8a_9}^{[3gv]}&=e^{2\eps\gamma_E}\int\frac{d^dk_1}{i\pi^{d/2}}\frac{d^dk_2}{i\pi^{d/2}}
\\
&\times
\frac{(k_1\cdot v_2)^{-a_7}(k_2\cdot v_3)^{-a_8}(k_2\cdot v_1)^{-a_9}}{(k_1^2)^{a_1}(k_2^2)^{a_2}((k_1+k_2)^2)^{a_3}(k_1\cdot v_1)^{a_4}(k_2\cdot v_2)^{a_5}(-(k_1+k_2)\cdot v_3 - 1)^{a_6}}. \nn
\end{align}
Using \path{Kira}~\cite{Klappert:2020nbg} we identify 16 basis integrals, 4 of which are in the top sector $I_{111111000}^{[3gv]}$. Differential equations are derived with respect to the three $x_{ij}$ variables. The rotation to canonical form \cite{Henn:2013pwa} is found using a combination of techniques. For the lower sectors, using the automated package \path{CANONICA}~\cite{Meyer:2017joq}, is enough.  
For the top sector we use \path{DlogBasis}~\cite{Henn:2020lye} to reveal uniform weight integrals.
In this way we find the following three integrals,
\begin{subequations}\begin{align}
  g_{13}^{[3gv]} &= \eps^4\frac{1-x_{13}^2}{x_{13}}\,\, I_{111111-100}^{[3gv]}\label{eq:g13}\,,\\
  g_{14}^{[3gv]} &= \eps^4\frac{1-x_{12}^2}{x_{12}}\,\, I_{1111110-10}^{[3gv]}\label{eq:g14}\,,\\
  g_{15}^{[3gv]} &= \eps^4\frac{1-x_{23}^2}{x_{23}}\,\, I_{11111100-1}^{[3gv]}\label{eq:g15}\,.
\end{align}\end{subequations}
The last required integral is found using \path{INITIAL}~\cite{Dlapa:2020cwj}
\begin{align}
  g_{16}^{[3gv]} &= \eps^3\,\left(2I_{111112-10-1}^{[3gv]} + (v_1\cdot v_3)\, I_{111112-100}^{[3gv]} + (v_2\cdot v_3)\, I_{11111200-1}^{[3gv]}\right) + \text{lower sectors}\label{eq:g16}.
\end{align}
Using this integral basis, we find a differential equation in the canonical form \cite{Henn:2013pwa}
\begin{equation}
\text{d}\,\mathbf{g}^{[3gv]} = \eps\sum_{\ell\in\mathcal{A}^{[3gv]}} c_\ell^{[3gv]}\,\text{d}\log(\ell)\,\mathbf{g}^{[3gv]}\,,\label{eq:DEsystem}
\end{equation}
where the set of $\ell$, the alphabet $\mathcal{A}$, is
\begin{align}\label{alphabettwoloops}
\mathcal{A}^{[3gv]}&=\{1-x_{12},\,x_{12},\,1+x_{12},\,1-x_{13},\,x_{13},\,1+x_{13},\,1-x_{23},\,x_{23},\,1+x_{23},
\\&\hskip2.5cm
x_{12} x_{13}+x_{23},\,x_{12} x_{23}+x_{13},\,x_{12}+x_{13} x_{23},\,1+x_{12} x_{13} x_{23}\}\,.
\nn
\end{align}
To solve the system in eq.~(\ref{eq:DEsystem}) we need a boundary condition. We choose $x_{ij} = 1$ for all $i$ and $j$. At this boundary point, the lower sector integrals are just iterated eikonal bubbles which evaluate to
\begin{align}
\int\frac{d^dk}{i\pi^{d/2}}\frac{1}{(k^2)^\a(k\cdot v - 1)^\beta} &= (-1)^{\alpha+\beta} \frac{2^{d-2 \a} \Gamma (d/2-\a) \Gamma (2 \a+\beta-d)}{\Gamma (\a) \Gamma (\beta)}, \label{eq:EikDef}
\end{align}
or, because of the differing regulators on different lines,
\begin{align}\label{eq:NewDef}
\hskip-0.5mm
\int\frac{d^dk}{i\pi^{d/2}}&\frac1{(k^2)^\a(k\cdot v - 1)^\beta(k\cdot v)^\gamma} \\&=  (-1)^{\alpha+\beta+\gamma}   \frac{2^{d-2\a}
\Gamma (d/2-\a) \Gamma (d-2 \alpha -\gamma) \Gamma (2 \alpha +\beta +\gamma -d)}{\Gamma (\alpha ) \Gamma (\beta ) \Gamma (d-2 \alpha )}.\nn
\end{align}
The top sector integrals $g_{13}^{[3gv]}$ to $g_{15}^{[3gv]}$ in eqs.~(\ref{eq:g13})-(\ref{eq:g15}) clearly vanish at such a boundary. However, $g_{16}^{[3gv]}$ in eq.~(\ref{eq:g16}) does not and depends on integrals more complicated than the bubble integrals in eqs.~(\ref{eq:EikDef}) and~(\ref{eq:NewDef}). 
Instead of attempting to compute them or determining from physical consistency conditions obtained from the differential equations (see e.g. \cite{Henn:2020lye}), we will fix the remaining boundary value from knowledge of the lightlike limit.

Solving the differential equation gives the following expansion in $\ep$ for $w_{3,\, 3gv}^{(2)}$ in eq.~(\ref{def:w3gvDef})
\begin{subequations}\begin{align}
    \web{2}{-3}_{3,\, 3gv}
    &=
    \mathcal{T}_{123}
    \frac{c_{0}}{16\pi^2}
    \left( r(x_{13})M_{000}(x_{13}) - r(x_{23})M_{000}(x_{23}) \right)
    \\
    \web{2}{-2}_{3,\, 3gv}
    &=
    \mathcal{T}_{123}
    \frac{c_{1}}{16\pi^2}
    \left( r(x_{13})M_{000}(x_{13}) - r(x_{23})M_{000}(x_{23}) \right)
    \\
    \web{2}{-1}_{3,\, 3gv}
    &=
    \frac{\mathcal{T}_{123}}{32\pi^2}
    \bigg\{
    \frac12\sum_{ijk}\ep_{ijk}r(x_{ij})M_{000}(x_{ij})M_{000}(x_{ik})^2
    \\\nn&\hskip3cm- c_2\left[r(x_{13})M_{000}(x_{13}) - r(x_{23})M_{000}(x_{23})\right]
    \bigg\}\,,
\end{align}
\end{subequations}
where the $c_i$ are constants coming from the remaining integral and the subscript denotes its transcendental weight. The web only has a single pole divergence, thus $c_0=c_1=0$. Together with eq.~(\ref{eq:Gamma121}) the complete three-line two-loop soft anomalous dimension is
\begin{align}
    \Gamma_3^{(2)}
    =
    -\mathcal{T}_{123}\frac1{8\pi^2}
    &\Bigg\{
    \sum_{ijk}\ep_{ijk}
    r(x_{ij})M_{000}(x_{ij})
    \left(
    r(x_{ik})M_{100}(x_{ik}) + \frac12 M_{000}(x_{ik})^2
    \right)\nn
    \\&\hskip1.7cm
    - c_2 \bigg(r(x_{13})M_{000}(x_{13}) - r(x_{23})M_{000}(x_{23})\bigg)
    \Bigg\}.
\end{align}
In the lightlike limit, when all $x_{ij}\to 0$ simultaneously, we have
\begin{align}
    \lim_{x_{ij}\to 0}
    \Gamma_{3}^{(2)}
    =
    \mathcal{T}_{123}\frac{c_2}{4\pi^2}\log\frac{x_{13}}{x_{23}}.
\end{align}
It is known that there is no tripole colour structure when all Wilson lines are lightlike, since we cannot construct conformally invariant cross ratios~\cite{Dixon:2009ur}. Therefore we conclude that $c_2=0$, giving
\begin{align}
    \Gamma_3^{(2)}
    =
    -\mathcal{T}_{123}\frac1{8\pi^2}
    \sum_{ijk}\ep_{ijk}
    r(x_{ij})M_{000}(x_{ij})
    \left(
    r(x_{ik})M_{100}(x_{ik}) + \frac12 M_{000}(x_{ik})^2
    \right),
\end{align}
in agreement with previous computations~\cite{Mitov:2009sv,Ferroglia:2009ep,Mitov:2010xw,Chien:2011wz}.

In summary, we see that
the calculation of the two-loop soft anomalous dimension matrix is greatly simplified 
by regulating only one of the multiple Wilson lines.
In the next section, we show that the new method is promising for higher-loop applications as well. 

\section{Three-loop calculation of a three-line web function}

\label{sec:threeloop}

\begin{figure}[t]
    \centering
    \includegraphics[width=0.5\textwidth]{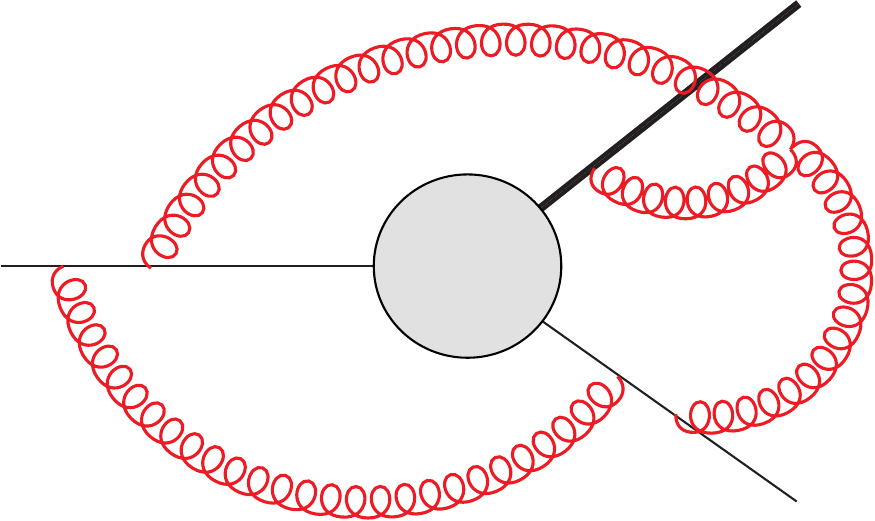}
    \caption{The three-loop $w_{122}^{(3)}$ diagram}
    \label{fig:w1223}
\end{figure}

To illustrate the potential of the new method, we evaluate in this section a three-loop web diagram.
This is a contribution to the soft anomalous dimension matrix of three Wilson lines at the three-loop order. 
Naturally, many further diagrams are required for the full calculation. Here, our intention is to provide a proof-of-concept by calculating a diagram, shown in Fig.~\ref{fig:w1223}, of genuine complexity.

The kinematic part of the integral shown in Fig.~\ref{fig:w1223} can be conveniently written as
\begin{align}
    \mathcal{F}_{122}^{(3)}=&\int\frac{d^dk_1}{i\pi^{d/2}}\frac{d^dk_2}{i\pi^{d/2}}\frac{d^dk_3}{i\pi^{d/2}}\\
    \times&\int d^dk_4
    \frac{\delta^{(d)}(k_1+k_2+k_3)\sum_{i,j,k}\eps_{ijk}(v_i\cdot v_j)( v_k\cdot k_i)}{k_1^2 k_2^2 k_3^2k_4^2(k_1\cdot v_1)((k_1-k_4)\cdot v_1)(k_4\cdot v_2)((k_2+k_4)\cdot v_2)(k_3\cdot v_3 -1)}.\nn
\end{align}
Using \path{Kira}~\cite{Klappert:2020nbg} we find that there are 66 basis integrals, belonging to 22 sectors. The dimension of the sector with the most number of integrals is 8. We then find the differential equation satisfied by these integrals.

For the transformation of the differential equation to canonical form, we found it convenient to transform sector-by-sector. The diagonal block transformations were found using either \path{CANONICA}~\cite{Meyer:2017joq} or using \path{DlogBasis}~\cite{Henn:2020lye} to identify one uniform weight integral and then using \path{INITIAL}~\cite{Dlapa:2020cwj} for the transformation. 

Once all the diagonal blocks are in canonical form, we then need to transform all the off-diagonal blocks. For this we used a combination of \path{CANONICA} and \path{Libra}~\cite{Lee:2020zfb}. We write our equation as
\begin{equation}
\text{d}\,\mathbf{g}^{[122]} = \eps\sum_{\ell\in\mathcal{A}^{[122]}} c_\ell^{[122]}\,\text{d}\log(\ell)\,\mathbf{g}^{[122]}\,,\label{eq:DEsystem122}
\end{equation}
where $c_\ell^{[122]}$ are $66\times 66$ constant matrices. The canonical differential equation matrices $c^{[122]}$, as well as the definition of the canonical basis $\mathbf{g}^{[122]}$, are included in the ancillary files. 
The alphabet appearing in the differential equation involves two additional letters compared to the two-loop case of eq. (\ref{alphabettwoloops}), namely
\begin{align}
    \{
 x_{13} + x_{12} x_{23} + x_{12} x_{13}^2 x_{23} + x_{13} x_{23}^2,
 x_{12} x_{13} + x_{23} + x_{13}^2 x_{23} + x_{12} x_{13} x_{23}^2
 \}\,.
\end{align}
In order to fully solve the differential equation we need a boundary condition. 
In practice, we consider kinematic points $v_i=v, v_j = v, v_k = -v$, where 
36 of the 66 integrals can be found analytically. Either they vanish or reduce to real-valued iterated integrals of the form given in eqs.~(\ref{eq:EikDef}) and~(\ref{eq:NewDef}). 
This does not completely fix the required constants for the single pole of $\mathcal{F}_{122}^{(3)}$ which we label as $\mathcal{F}_{122}^{(3,-1)}$. 
We can find the remaining constants by imposing finiteness of the differential equation on single-variable slices. Consider the single-variable regime $(x_{12}, x_{13}, x_{23}) = (1,x,x)$. Then we impose the condition that eq.~(\ref{eq:DEsystem122}) remains finite as $x\to 1$ and $x\to\pm i$. Doing this also for the other permutations we find we can fix all the boundary conditions.

Thus we arrive at the result
\begin{align}\label{eq:w122Res}\nn
    \mathcal{F}_{122}^{(3,-1)}
    =
    r(x_{12})^2&f_A(x_{12},x_{13},x_{23})
    +
    r(x_{12})r(x_{13}) f_B(x_{12},x_{13},x_{23})\\&
    +
    r(x_{12})r(x_{23}) f_C(x_{12},x_{13},x_{23}),
\end{align}
where the functions $f_A$, $f_B$ and $f_C$ are of pure weight five. They can be written in terms of classical polylogarithms $\text{Li}_n(-s)$ up to weight $n=5$, with the following arguments,
\begin{align}
s \in \{ x_{12}, -x_{12},  x_{13}, -x_{13}, \frac{x_{12}}{x_{13} x_{23}} , \frac{x_{12} x_{13}}{x_{23}}, \frac{ x_{12} x_{23}}{x_{13}} ,  x_{12} x_{13} x_{23} \}.
\end{align}
Their symbols are drawn from a smaller set of alphabets given in eq.~(\ref{alphabettwoloops}).
The expressions for $f_{A,B,C}$ are given explicitly in the ancillary files.

We would like to present the total contribution of web function $w^{(3)}_{122}$ after symmetrization  under permutations of Wilson lines.
Hence let us introduce a basis  $\{  g_A^\pm , g_B^\pm , g^\pm_C \}$ defined as follows
\begin{align}
g_A^\pm (3,2,1)  &= \pm  g_A^\pm (3,1,2)  \equiv  \frac12  \left[  f_A (x_{12},x_{13},x_{23}) \pm  ( x_{13} \leftrightarrow  x_{23} ) \right]  , \nn \\ 
g_B^\pm  (3,2,1)  &= \pm g_B^\pm (2,3,1)  \equiv  \frac12  \left[  f_B(x_{12},x_{13},x_{23}) \pm  ( x_{12} \leftrightarrow  x_{13} ) \right] , \nn \\ 
g_C^\pm (3,2,1)  &=  \pm g_C^\pm (1,2,3)  \equiv  \frac12  \left[  f_C(x_{12},x_{13},x_{23})  \pm ( x_{12} \leftrightarrow  x_{23} ) \right] \,.  
\end{align} 
where we adopt a short-hand notation $g(i,j,k) = g(\alpha_i, \alpha_j , \alpha_k) $ where 
\begin{align}
\alpha_3 \equiv x_{12}\, , \quad \alpha_2 \equiv x_{13}\,, \quad  \alpha_1 \equiv x_{23} \,.
\end{align}
In this notation the permutation among Wilson lines $v_{i,j,k}$ is equivalent to the permutation among arguments of the function $g(i,j,k)$.  

Summing over all six permutations of the web diagram  $\cF^{(3)}_{122}$, we obtain 
\begin{align}\label{F122Colorsum}
& \mathcal{C}(1,2,3) \, \cF_{122}^{(3)} (3,2,1) + {\rm {perm.} } = \frac1\epsilon \times    \\
& \Bigg\{ r (1)^2  \,  \left(  [\mathcal{C} (3,2,1) + \mathcal{C} (2,3,1)] \, g_A^+ (1,2, 3)  + [ \mathcal{C} (3,2,1)- \mathcal{C} (2,3,1)] \, g_A^- (1 , 2, 3) \right)    \nn \\
& + r (3) r (2)  \,  \left(  [\mathcal{C} (1,2,3)+ \mathcal{C} (1,3,2) ] \, g_B^+ (3,2 , 1)  + [\mathcal{C} (1,2,3)- \mathcal{C} (1,3,2)] \, g_B^- (3 , 2, 1) \right.    \nn \\
&\qquad \qquad \; \; +  \left.  [\mathcal{C} (2,1,3)+ \mathcal{C} (3,1,2) ] \, g_C^+ (3, 1, 2)  + [\mathcal{C} (2,1,3) - \mathcal{C} (3,1,2)] \, g_C^- (3,1,2) \right)     \nn \\
&+ {\rm cyc.}  \Bigg\} .  \nn
\end{align}
The expression in the braces is manifestly symmetric under exchange $ 2 \leftrightarrow 3\,$, and cyc. represents the sum of two cyclic permutations. The color factor $\mathcal{C}(1,2,3) =   \T_1^c \,  [\T_3 \cdot \T_1, \T_3 \cdot \T_2 ] \, \T_2^c $
 is associated with the diagram we computed, whose maximally non-abelian component can be decomposed onto the following basis, 
\begin{align}
&   \T_1^c \,  [\T_3 \cdot \T_1, \T_3 \cdot \T_2 ] \, \T_2^c  - \frac12  \Big\{ \T_1 \cdot \T_2  ,   [\T_3 \cdot \T_1, \T_3 \cdot \T_2 ]   \Big\}  =  \\
& \frac{1}{4} \,   [\T_3^a, \T_3^b]  \, [\T_1^c, \T_1^a]  \,  \{ \T_2^b , \T_2^c \}    
+\frac14 \,  [\T_3^a, \T_3^b]  \,\{  \T_1^c , \T_1^a  \} \, [ \T_2^b,  \T_2^c]  + \frac12 \,  [\T_3^a, \T_3^b]  \, [\T_1^c, \T_1^a]  \,  [ \T_2^b , \T_2^c ] \, .
  \nn 
\end{align} 

We would like to isolate the contributions to $F_1-$ and $F_2-$term in the soft anomalous dimension. For this purpose  it is convenient to regroup terms associated with $r(3) r(2)$  in the following way 
\begin{align}
&  r (3) r (2)  \,\big(  [\mathcal{C} (1,2,3)+\mathcal{C} (2,1,3) ] \, g_{BB}  + [\mathcal{C} (1,3,2) + \mathcal{C} (3,1,2) ] \, g_{CC}      \nn \\
& \qquad \qquad +  \left.  [\mathcal{C} (1,2,3)-\mathcal{C} (2,1,3)  + \mathcal{C} (1,3,2)  -  \mathcal{C} (3,1,2) ] \, g_{+-}   \right. \nn \\
&  \qquad \qquad +  [\mathcal{C} (1,2,3)-\mathcal{C} (2,1,3)  - \mathcal{C} (1,3,2) + \mathcal{C} (3,1,2) ]  \, g_{--}       \big) , 
\end{align} 
where
\begin{align}
&g_{BB} = g_B^+ (3,2,1) + g_C^+ (3,1,2) + g_B^- (3,2,1) + g_C^- (3,1,2) ,  \quad g_{CC}= g_{BB}|_{ 2\leftrightarrow 3} \nn \\
& g_{+-} =  g_B^+ (3,2,1) - g_C^+ (3,1,2), \quad  g_{--} =  g_B^- (3,2,1) - g_C^- (3,1,2) .
\end{align}
Investigating the structure of color factors,  we observe that only $\{  g_{A}^-, g_{--}\}$ contribute to the $F_1-$term in the soft anomalous dimension , whereas only $\{ g_{A}^+, g_{BB}, g_{CC} \}$ contribute to the $F_2-$term.  The expressions for $ \{g_{A}^-(1,2,3), g_{--} , g_{A}^+(1,2,3), g_{BB} \} $ are given in the following, which involve  products of logarithms with classical polylogarithms up to weight four.
\begin{align} 
g_{A}^-& =  -\frac{8}{9}\,  l_{+} l_{-} h_3(1)  , \label{gAminus}
\\
g_{--} &= - \frac49 \, \bigg[ -6 \, l_{-}  \cL_4^{c} + 6 \, l_{+}  \cL_4^d +4\,  l_3  h^b_4 (2)  - 4\, l_2 h^b_4(3)    \label{gmminus}
\\
& 
-3 \, l_{-} l_{+} \cL_3^b  +  6\, l_1 l_{-} \cL_3^c  -6 \, l_1  l_{+} \cL_3^d \nn \\
& +3\, l_1 l_{-} l_{+} \cL_2^b  -  l_{-} (3\, l_1^2 + \pi^2 ) \cL_2^c +  l_{+}(3 \,l_1^2 + \pi^2) \cL_2^d  \nn \\
& 
+ l_2 l_3  l_-  l_{+} \cL_1^a  +  
 \frac12 \, l_{-}  l_{+}  (-3 \, l_1^2 + l_2^2 + l_3^2 + \pi^2 ) \cL_1^b  
 +l_1 l_{-}  (l_1^2 +\pi^2) \cL_1^c  -l_1 l_{+}  (l_1^2 + \pi^2) \cL_1^d  \bigg] , \nn \\
g_{A}^+& = \frac{4}{9}  \, l_1 \bigg[  
- 6 \, \cL_4^a 
  - 6 \, l_{+} \cL_3^{c} - 6  \, l_- \cL_3^d     - 8 \, h^a_4(1) 
  \label{gAplus} \\
& -\, (-3\, l_1^2 + 3\,  l_2^2 + 3\, l_3^2 + \pi ^2) \cL_2^a 
-6 \, l_2 l_3 \cL_2^b
    \nn \\
  & -2\,  l_1 (l_1^2 + \pi^2) \cL_1^a  
  -   l_+  (-3\, l_1^2 + l_+^2 +\pi^2 )  \cL_1^{c} 
 - l_-  (-3\, l_1^2 + l_-^2 + \pi^2 ) \cL_1^{d}  \bigg]  , \nn \\
g_{BB}& = 
-\frac{4}{9}\,l_3 \bigg[ -6 \, (\cL_4^{c}+\cL_4^{d})  -6 \, l_3 \cL_3^{b} + 6\, l_1 (\cL_3^{c}+\cL_3^{d} )  + 8\, h_4^a (2)   \label{gBB} \\
 & +6\, l_1 l_3 \cL_2^{b} + (-3 \, l_1^2 + 3\, l_{+} l_{-}  - \pi^2) (\cL_2^{c}+ \cL_2^d) \nn \\
&  +2\, l_2 (l_2^2 + \pi^2) \cL_1^a  - 
 l_3 (3 \,l_1^2 - 3\,l_2^2 + l_3^2 + \pi^2) \cL_1^b + 
 l_1 (l_1^2 - 3 \, l_{+} l_{-} + \pi^2) (\cL_1^c +\cL_1^d) 
  \bigg] . \nn   
\end{align}
Here we define logarithms
\begin{align}\label{logdef}
l_1 &= \ln \alpha_1 ,\quad   l_2 = \ln \alpha_2 ,\quad  l_3= \ln \alpha_3 ,  \quad  l_{+} =  \ln \alpha_2 \alpha_3 , \quad l_{-} =  \ln \frac{\alpha_2}{\alpha_3},  
\end{align}
and basis of polylogarithms 
\begin{align} \label{Lndef}
\cL_{n}^a& = \frac12 \left[  \text{Li}_n \left(- \frac{\alpha_1}{\alpha_2 \alpha_3} \right)+\text{Li}_n \left(- \frac{\alpha_1 \alpha_3}{ \alpha_2} \right)+ \text{Li}_n \left(- \frac{\alpha_1 \alpha_2}{ \alpha_3} \right)+\text{Li}_n \left(-  \alpha_1 \alpha_2 \alpha_3 \right) \, \right]    \; \pm  \left(\alpha \leftrightarrow \frac1\alpha  \right)  \nn \\
\cL_{n}^b& = \frac12 \left[ \text{Li}_n \left(- \frac{\alpha_1}{\alpha_2 \alpha_3} \right)-\text{Li}_n \left(- \frac{\alpha_1 \alpha_3}{ \alpha_2} \right)- \text{Li}_n \left(- \frac{\alpha_1 \alpha_2}{ \alpha_3} \right)+\text{Li}_n \left(-  \alpha_1 \alpha_2 \alpha_3\right)  \, \right]  \;   \pm  \left(\alpha \leftrightarrow \frac1\alpha  \right)  \nn \\
\cL_{n}^c& = \frac12 \left[ \text{Li}_n \left(- \frac{\alpha_1}{\alpha_2 \alpha_3} \right)- \text{Li}_n (-  \alpha_1 \alpha_2 \alpha_3 )  \, \right]  \;   \pm  \left(\alpha \leftrightarrow \frac1\alpha  \right) \nn \\
\cL_{n}^d& = \frac12 \left[ \text{Li}_n \left(- \frac{\alpha_1 \alpha_3}{ \alpha_2} \right)-  \text{Li}_n \left(- \frac{\alpha_1 \alpha_2}{ \alpha_3} \right)  \, \right] \; \pm\left(\alpha \leftrightarrow \frac1\alpha  \right)\,  ,
\end{align}
where $\pm$ stands for $+$ if $n$ is odd, and $-$ if $n$ is even.  By construction, these basis are (anti-)symmetric under the transformation which inverts all $\alpha$'s.

In addition, there are terms which depend on a single cusp angle. We group them together and introduce the following set of functions  
\begin{align} \label{hdef2}
h_3(i) & =   - 24\,  li_3 (i),    
\nn \\
h^a_4(i) &=  6\,   li_4 (i)  -4 \, l_i\, li_3(i) + ( l_i^2+ \pi^2)  \, li_2(i) \,, \quad  
 h^b_4(i)  =  72 \, li_4 (i) +  h_4^a (i) ,
\end{align}
where 
\begin{align}\label{lindef}
li_{n} (i) \equiv \frac1{2^n}  \left[ \text{Li}_n (\alpha_i^2)  - \zeta_{n}
-\frac{ 2 }{n-1} \ln \alpha_i \, ( \text{Li}_{n-1} (\alpha_i^2)  - \zeta_{n-1}) \right]  \;  \pm  \left(\alpha_i \leftrightarrow \frac{1}{\alpha_i}\right) \,.
\end{align}
Again, $\pm$ stands for $+$ if $n$ is odd, and $-$ if $n$ is even, $ \forall n \in \{ 2,3,4\}$. In the special case where $n=2$, the $\zeta_{n-1}$ term in \eqref{lindef} should be set to 0.

Let us emphasize that the basis of polylogarithms given  in \eqref{logdef},\eqref{Lndef} and  \eqref{lindef} satisfy the following properties. Firstly, they are single-valued in the so-called \textit{Euclidean region}, where $\alpha_1, \alpha_2, \alpha_3 >0$. 
In particular the branch cuts on the positive real axis of $\alpha_i$ cancel  in \eqref{lindef}. 
Secondly, each member in the set of function we introduced, i.e.  
\[\{ l_1, l_2, l_3,  \cL_{n}^a, \cL_{n}^b, \cL_{n}^c \pm \cL_{n}^d , li_n \} \]
is either symmetric or anti-symmetric under the transformation which inverts any set of cusp angle variables $(\alpha_i \rightarrow \alpha_i^{-1},\, \forall\, i )$.  
Moreover, the functions 
\[\{ l_{+}, l_{-}, \cL_{n}^a, \cL_{n}^b, \cL_{n}^c, \cL_{n}^d \} \]
are (anti-)symmetric under the exchange $(2 \leftrightarrow 3)$. 
Given these nice properties, the form of the final result is highly constrained.

\section{Discussion and outlook}
\label{sec:outlook}

In this paper we proposed and tested a new regularization scheme for computing the soft anomalous dimension matrix.
We considered the ratio between the correlation function for $n$ and $n-1$ Wilson-line operators,  which is free from IR divergences, provided that 
the $n-$th Wilson-line is dressed with an offshellness regulator.  
Based on this observation, we defined a renormalized quantity $\cS_{0,\cdots,1}^{\text{fin.}}$ 
that satisfies the renormalization group equation  
\begin{align}
 \left[ \mu \frac{\partial}{\partial \mu } + \beta(\alpha) \frac{\partial}{\partial \alpha} \right]
 \mathcal{S}_{ 0,\cdots,1}^{\text{fin.}} = {\bf \Gamma_{n-1}} \,\mathcal{S}_{0,\cdots, 1}^{\text{fin.}} - \mathcal{S}_{0,\cdots, 1}^{\text{fin.}} \, {\bf{\Gamma_n} }   \,.   
\end{align} 
This allows us to determine the  
soft anomalous dimension matrix for an arbitrary number of massive particles, by sequentially computing $\mathcal{S}_{0,1}, \mathcal{S}_{0,0,1},\cdots, \mathcal{S}_{0,\cdots,1}$ and extracting their UV poles.

We tested the novel method by demonstrating how the known one- and two-loop results are obtained within this setup. Employing the differential equations method, we note that calculations are simpler compared to the traditional setup where all Wilson lines are regulated in the IR.
As a proof of principle, we evaluated one of the three-loop web diagrams that contributes to the three-Wilson-line soft anomalous dimension matrix ${\bf\Gamma}_3$. A natural next step, beyond the scope of this paper, is to evaluate the other relevant Feynman diagrams contributing to this observable.

Our two-loop and sample three-loop calculations shed light on the new features that might appear in ${\bf\Gamma}_3$.
The final formula for the two-loop soft anomalous dimension matrix depends on the following alphabet only,
\begin{align}\label{alphabettwoloopsfinal}
\{1-x_{12},\,x_{12},\,1+x_{12},\,1-x_{13},\,x_{13},\,1+x_{13},\,1-x_{23},\,x_{23},\,1+x_{23}\}\,.
\end{align}
Although this depends on three variables, the dependence is factorized, which implies that any function within this alphabet is a sum of products of single-variable functions.
In-contrast, the web diagrams we computed involve the thirteen-letter alphabet (\ref{alphabettwoloops}). Compared to \eqref{alphabettwoloopsfinal}, this involves the following new alphabet letters,
\begin{align}
    \{ x_{12}+ x_{13} x_{23}, x_{13}+ x_{12} x_{23},x_{23}+ x_{12} x_{13}, 1+ x_{12}x_{13} x_{23} \}\,.
\end{align} 
This implies more complicated and richer structures for the transcendental functions.\footnote{Also, the result of ref. \cite{Liu:2022elt} for the soft anomalous dimension matrix for one massive and two massless particles may suggest the appearance of further alphabet letters.}
(At the same time, it is interesting that $\cF_{122}$  is given by products of logarithms with classical polylogarithms whose arguments are drawn from a small set of variables.)
These observations motivate a more systematic study on the function space of multi-line soft anomalous dimension,
which will be valuable input for bootstrap approaches.

A further interesting direction is to investigate the web function in $\mathcal{N}=4$ super Yang-Mills (sYM).
At the three-loop order, the difference between the corresponding soft anomalous dimension matrices is captures by simpler, matter-dependent terms.
Interestingly, those terms have been observed to 
 exhibit an universal iterative structure \cite{Grozin:2015kna} in the two-line case, namely
\begin{align}
\Gamma_{\text{cusp}} (\phi ) = C_R \frac{a}{ \pi} \left[  \Omega(\phi) +   \frac{a}{ \pi}  C_A \Omega_A( \phi)  +  \left(\frac{a}{ \pi} \right)^2   C_{A}^2 \Omega_{AA}( \phi)  \right] + \cO(\alpha_s^4 )\,.
\end{align} 
The angle-dependent function $\Omega$'s are independent of the particle content of the gauge theory, whereas $n_f-$ and $n_s-$dependence can be associated with an effective coupling constant $a$.  
(This holds up to three loops; at four loops terms proportional to the quartic Casimir color component break this pattern \cite{Bruser:2020bsh}.)
It would be interesting to see whether such a pattern exists for the tripole function $F_1^{(2)}$ and $F_1^{(3)}$ as well. 
Finally, in addition to considering the standard Wilson line operator, it is possible  
to consider the (locally supersymmetric) Maldacena operator, which includes a term coupling to scalars. 
This leads to additional sets of angle parameters 
in the flavour space  of scalars that the correlation functions depend on, and that can be used to organize calculations and define novel limits \cite{Correa:2012nk}.

\section{Acknowledgements}
We would like to thank Einan Gardi for enlightening discussions. 
This research received funding from the European Research Council (ERC) under the European Union's Horizon 2020 research and innovation programme (grant agreement No 725110), {\it Novel structures in scattering amplitudes}. 
KY is supported by the National Natural Science Foundation of China under Grant No. 12357077 and would like to thank the sponsorship from Yangyang Development Fund.  CM has been partially supported by the Italian Ministry of University and Research (MUR) through grant PRIN20172LNEEZ and by Compagnia di San Paolo through grant TORPS1921EX-POST2101.

\bibliographystyle{JHEP}

\providecommand{\href}[2]{#2}\begingroup\raggedright\endgroup


\end{document}